\documentclass[twocolumn,english]{revtex4-2}
\usepackage[T1]{fontenc}
\usepackage[latin9]{inputenc}
\usepackage{xcolor}
\usepackage{amsmath}
\usepackage{amssymb}
\usepackage{graphicx}

\makeatletter

\newcommand*\LyXZeroWidthSpace{\hspace{0pt}}

\usepackage{xcolor}
\definecolor{olivegreen}{RGB}{85, 107, 47}
\usepackage[colorlinks,bookmarksopen,bookmarksnumbered,citecolor=red, linkcolor=red, urlcolor=red]{hyperref}
\usepackage{placeins}

\makeatother

\usepackage{babel}
\begin{document}
\title{Barrier induced stalemate-consensus transition of self-propelled participants
subject to majority rule}
\author{Yan-Wen Xiao}
\thanks{These authors contributed equally to this work.}
\affiliation{Institute for Theoretical Physics, School of Physics, South China
Normal University, Guangzhou 510006, China}
\affiliation{Key Laboratory of Atomic and Subatomic Structure and Quantum Control
(Ministry of Education), Guangdong Basic Research Center of Excellence
for Structure and Fundamental Interactions of Matter, School of Physics,
South China Normal University, Guangzhou 510006, China}
\affiliation{Guangdong Provincial Key Laboratory of Quantum Engineering and Quantum
Materials, Guangdong-Hong Kong Joint Laboratory of Quantum Matter,
South China Normal University, Guangzhou 510006, China}
\author{Wei-Chen Guo}
\thanks{These authors contributed equally to this work.}
\affiliation{Institute for Theoretical Physics, School of Physics, South China
Normal University, Guangzhou 510006, China}
\affiliation{Key Laboratory of Atomic and Subatomic Structure and Quantum Control
(Ministry of Education), Guangdong Basic Research Center of Excellence
for Structure and Fundamental Interactions of Matter, School of Physics,
South China Normal University, Guangzhou 510006, China}
\affiliation{Guangdong Provincial Key Laboratory of Quantum Engineering and Quantum
Materials, Guangdong-Hong Kong Joint Laboratory of Quantum Matter,
South China Normal University, Guangzhou 510006, China}
\author{Bao-Quan Ai}
\email{aibq@scnu.edu.cn}

\affiliation{Institute for Theoretical Physics, School of Physics, South China
Normal University, Guangzhou 510006, China}
\affiliation{Key Laboratory of Atomic and Subatomic Structure and Quantum Control
(Ministry of Education), Guangdong Basic Research Center of Excellence
for Structure and Fundamental Interactions of Matter, School of Physics,
South China Normal University, Guangzhou 510006, China}
\affiliation{Guangdong Provincial Key Laboratory of Quantum Engineering and Quantum
Materials, Guangdong-Hong Kong Joint Laboratory of Quantum Matter,
South China Normal University, Guangzhou 510006, China}
\author{Liang He}
\email{liang.he@scnu.edu.cn}

\affiliation{Institute for Theoretical Physics, School of Physics, South China
Normal University, Guangzhou 510006, China}
\affiliation{Key Laboratory of Atomic and Subatomic Structure and Quantum Control
(Ministry of Education), Guangdong Basic Research Center of Excellence
for Structure and Fundamental Interactions of Matter, School of Physics,
South China Normal University, Guangzhou 510006, China}
\affiliation{Guangdong Provincial Key Laboratory of Quantum Engineering and Quantum
Materials, Guangdong-Hong Kong Joint Laboratory of Quantum Matter,
South China Normal University, Guangzhou 510006, China}
\begin{abstract}
Natural or artificial barriers---such as the Himalayas, the Berlin
Wall, or the Korean Demilitarized Zone---can significantly impede
human migration. As a consequence, they may also hinder the dissemination
of opinions within society, thereby contributing to divergent geopolitical
landscapes and cultural developments. This raises a fundamental question:
how do such barriers influence the opinion dynamics of mobile agents,
such as human beings? In particular, can a barrier induce transitions
in collective opinion states among spatially segregated groups? Here,
we investigate the opinion dynamics governed by majority rule in a
minimal model comprising self-propelled agents with binary opinions
performing random walks within a closed space divided by a barrier.
We focus on the conditions under which initially segregated clusters
of agents with opposing opinions can reach consensus. Our results
reveal the existence of a critical barrier size that marks a transition
between stalemate and consensus states. Near this critical point,
the relaxation time to reach consensus from an initial stalemate exhibits
a power-law divergence. This barrier-induced stalemate-consensus transition
in a simple agent-based model offers new insights into the role of
physical or social barriers in shaping opinion dynamics and social
structures.
\end{abstract}
\maketitle

\section{Introduction}

Humans, as social beings, constantly exchange opinions, giving rise
to complex and intriguing opinion dynamics \citep{RevModPhys.81.591,RevModPhys.88.045006,PhysRevLett.110.088701,PhysRevLett.103.018701}.
By identifying fundamental principles and general mechanisms underlying
these seemingly elusive processes, theoretical research on opinion
dynamics has demonstrated its practical value in diverse areas, such
as predicting election outcomes \citep{galam_contrarian_2004,bernardes_election_2002},
assessing advertising effectiveness \citep{schulze_advertising_2003,sznajd-weron_how_2003},
planning transportation systems \citep{le_pira_simulating_2015},
and managing public opinion \citep{han_soft_2006,kurz_optimal_2015,ding_opinion_2016}.
However, due to the inherently subjective and abstract nature of opinions,
developing models with robust explanatory power for the rich phenomena
of opinion dynamics remains a persistent challenge.

A particularly captivating topic in this field is the confrontation
of binary opinions, such as those observed in two-party political
systems. Several well-known models have been proposed to address this
and related phenomena, including voter-like models \citep{richard_a_holley_ergodic_1975,lambiotte_dynamics_2009,crokidakis_inflexibility_2015,khalil_zealots_2018},
Ising-like models \citep{Chandra_PRE_2012,Galam_PRE_2015,Anteneodo_PRE_2017,devauchelle_dislike_2024},
the Sznajd model \citep{sznajd-weron_opinion_2000}, the Galam model
\citep{galam_minority_2002,ellero_stochastic_2013}, the Deffuant-Weisbuch
model \citep{Deffuant_Weisbuch_model,li_bounded-confidence_2023,huang_effects_2018},
and the Hegselmann-Krause model \citep{Hegselmann_Krause_model,han_opinion_2019}.
Beyond traditional lattice-based models, research has also extended
to models on networks \citep{holme_collective_2016,fu_coevolutionary_2008,wang_dual_2011,cao_mixed_2008,martins_mobility_2008,qu_nonconsensus_2014,benczik_opinion_2009,sugishita_opinion_2021,meng_opinion_2018,maity_opinion_2012,tartaglia_percolation_2015,bartolozzi_stochastic_2005}
and hypergraphs \citep{papanikolaou_consensus_2022,landry_opinion_2023,noonan_dynamics_2021},
where the structures in which opinion agents operate can be either
static or coevolve with the agents.\textcolor{black}{{} }

In most of these models, opinion agents are typically represented
as binary variables, with their evolution governed by specific rules
that reflect decision-making processes. These rules are often inspired
by physical analogies. For example, the majority rule and its variants
posit that agents tend to align their opinions with the dominant opinion
of their neighbors \citep{galam_minority_2002}, akin to the ferromagnetic
coupling in spin systems. However, two important aspects that are
common in real-world scenarios have been relatively overlooked. First,
the mobility of opinion agents can drive both conflict and integration,
potentially giving rise to richer phenomena, such as the scaling properties
of human travel \citep{brockmann_scaling_2006,Song2010}. Second,
physical geographical barriers, such as mountains or national borders,
can impede agent migration and influence regional opinion distribution,
as evidenced by sociocultural patterns \citep{wu_geographical_2018,xueqiong_tang_conceptualizing_2014}.
Thus, incorporating the mobility of opinion agents alongside physical
barriers could offer a more accurate and comprehensive framework for
capturing the rich phenomena of opinion dynamics.

\begin{widetext}

\begin{figure}[h]
\begin{centering}
\includegraphics[width=1\textwidth]{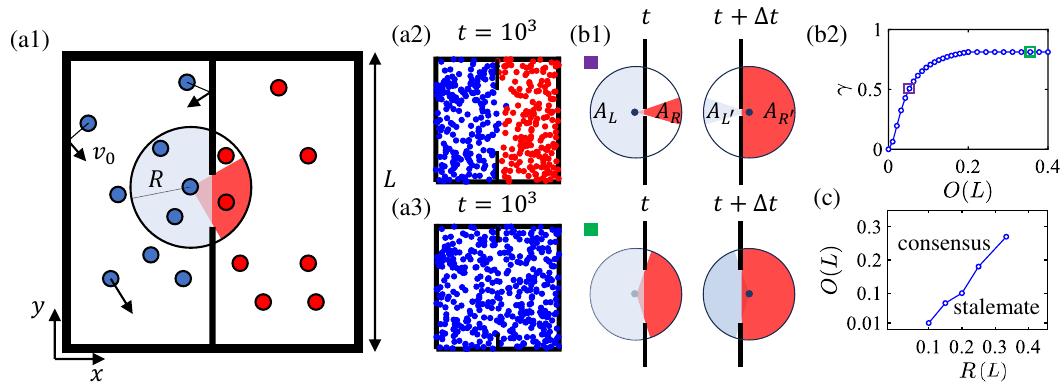}
\par\end{centering}
\caption{\protect\label{fig:1}(a1) Schematic illustration of the system. Particles
with different opinions are represented by circles of different colors.
Each particle performs a random walk at constant speed $v_{0}$, and
updates its opinion based on the majority opinion of particles within
its circular view field of radius $R$. Both the barrier and the boundary
reflect particles upon collision and block the view of each particle.
(a2,~a3) Representative configurations of the system with different
barrier size after evolving for the same duration ($t=10^{3}$). The
system with a small barrier reaches a consensus state, whereas the
system with a large barrier remains in stalemate. (b1) Illustration
of how a particle going across the opening experiences a change in
the surrounding opinion environment. Colored regions denote the visible
areas to this particle occupied by other particles with distinct opinions.
(b2) The dependence of opinion environment factor $\gamma$ on the
barrier opening $O$, corresponding to the dynamic process illustrated
in (b1). The colored square frames in (b2) qualitatively indicate
the corresponding opening sizes in (b1). (c) Phase diagram of the
system in terms of the view radius $R$ and the barrier opening $O$.
The markers indicate the critical opening size $O_{c}$, which separates
the final state of our system between consensus and stalemate.}
\end{figure}

\end{widetext}

In this work, we explore the coupled effects of agent mobility and
physical barriers on opinion dynamics by considering a minimal model
of self-propelled particles undergoing random walks in a two-dimensional
enclosed square domain with a central barrier {[}cf. Fig.~\ref{fig:1}(a1){]}.
At each time step, particles update their positions via random motion
and then synchronously adopt the majority opinion of those within
their field of view. We focus on the conditions under which initially
segregated clusters of agents with opposing opinions can reach consensus.
We find that the system exhibits a transition from a stalemate to
a consensus state at a critical barrier opening size $O_{c}$. Near
this critical point, the relaxation time required to reach consensus
from an initial stalemate diverges following a power-law scaling.
These findings highlight a novel barrier-induced stalemate-consensus
transition and offer new insights into the role of physical obstacles
in shaping the collective opinion dynamics of mobile agents.

\section{System and model\protect\label{sec:System-and-model}}

To investigate the influence of agent mobility and physical barriers
on opinion dynamics, we consider a minimal system consisting of $N$
self-propelled agents (particles) with binary opinions, performing
random walks in a 2D square box of size $L$ {[}cf. Fig.~\ref{fig:1}a(1){]}.
At the center of the box, a line-shaped barrier with an opening of
length $O$ represents real-world geographical or artificial obstacles,
such as mountains, national borders, or modern information cocoons
created by recommendation algorithms, which hinder the exchange of
opinions. The particles elastically collide with either the barrier
or the box boundary when they approach these boundaries. Each particle
has a circular field of view with radius $R$, and the presence of
the barrier may obstruct part of its view. A particle perceives the
opinions $s_{i}$ of other particles within its field of view and
updates its own opinion at each time step according to the simple
majority rule, as described by Eq.~(\ref{eq:Majority_rule}). The
equations of motion for the system assume the following discrete form:
\begin{align}
\boldsymbol{r}_{j}(t+\Delta t) & =\boldsymbol{r}_{j}(t)+\boldsymbol{v}_{j}(t)\Delta t,\label{eq:2}\\
s_{j}(t+\Delta t) & =\begin{cases}
s_{j}(t) & \text{if \ensuremath{\sum_{i\in V_{j}}s_{i}s_{j}>0},}\\
-s_{j}(t) & \text{otherwise,}
\end{cases}\label{eq:Majority_rule}
\end{align}
where $\boldsymbol{r}_{j}(t)$, $\boldsymbol{v}_{j}(t)$, and $s_{j}(t)$
represent the position, velocity, and binary opinion of the $j$-th
particle at time $t$, respectively, and $\Delta t$ is the time step.
The velocity of the particle $\boldsymbol{v}_{j}$ is given by ${\color{blue}{\color{gray}{\color{black}\boldsymbol{v}_{j}=v_{0}\cos(\theta_{j}(t))\boldsymbol{\hat{e}_{x}}+v_{0}\sin(\theta_{j}(t))\boldsymbol{\hat{e}_{y}}}}}$,
where $v_{0}$ is a constant speed and $\theta_{j}(t)$ is a random
variable uniformly distributed within the interval $[0,2\pi]$. The
opinion $s_{j}(t)$ takes the values $\pm1$, representing two opposing
opinions, and $V_{j}$ denotes the set of particles within the field
of view of the $j$-th particle.

Our model differs fundamentally from classical opinion dynamics frameworks.
Unlike the Galam model \citep{galam_minority_2002}, where agents
interact in randomly assigned groups, or the voter \citep{richard_a_holley_ergodic_1975}
and Sznajd \citep{sznajd-weron_opinion_2000} models defined on static
networks, here agents have finite-range interactions within dynamically
evolving neighborhoods due to limited view field and mobility of themselves.

Agent mobility in our model creates a dynamic tension: While spatial
movement facilitates opinion spreading, the barrier suppresses opinion
exchange. This competition leads to a fundamental question---what
phase transition arises in such a system? As we show, the system exhibits
a barrier-induced stalemate-to-consensus transition, with the consensus
timescale diverging algebraically at the critical point.

\begin{widetext}

\begin{figure}[h]
\begin{centering}
\includegraphics[width=0.75\textwidth]{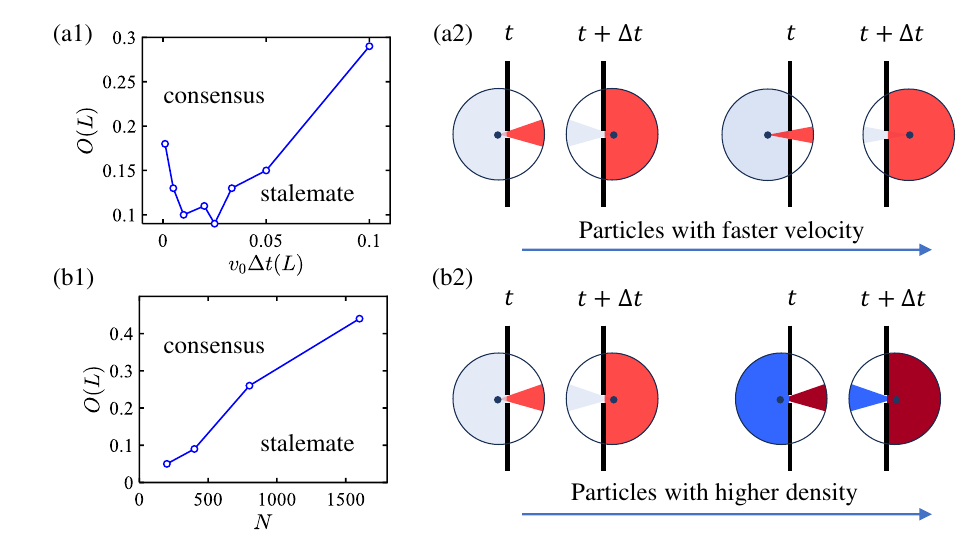}
\par\end{centering}
\caption{\textcolor{violet}{\protect\label{fig:2}}(a1, b1) Influence of particle
velocity and particle density on the critical opening size, respectively.
In (a1), the minimal $O_{c}$ is observed when the particle velocity
$v_{0}\Delta t=0.025L$. The effect of particle velocity on $O_{c}$
can be understood as follows: Particles with higher velocity are more
likely to penetrate deeper into the adversarial particle cluster,
thereby suppressing $\gamma$. This suppression makes it more likely
for a particle to flip its opinion at $t+\Delta t$, explaining the
monotonic increase of $O_{c}$ with particle velocity {[}see (a2){]}.
In (b1) the influence of particle density on $O_{c}$ is also significant.
Increased particle density reduces relative density fluctuations within
a particle's view field, thereby suppressing $\gamma$ (to prevent
it from exceeding $1$). As a result, a particle is more likely to
switch its opinion at $t+\Delta t$ {[}see (b2){]}.}
\end{figure}

\end{widetext}

\section{Barrier-induced stalemate-consensus transition and its critical scaling\protect\label{sec:results-and-discussion}}

Most of the results presented below are obtained through direct numerical
integration of the system's equations of motion, Eqs.~(\ref{eq:2},~\ref{eq:Majority_rule}).
We set $L=1$ and $\Delta t=1$ as the characteristic units of length
and time, respectively. Unless otherwise specified, the system comprises
$N=400$ particles with an interaction range defined by a field-of-view
radius $R=L/5$. At initial time ($t=0$), the particles are uniformly
distributed on opposite sides of the barrier: $N/2$ particles with
opinion $-1$ are placed on the left, while the remaining $N/2$ particles
with opinion $+1$ are positioned on the right, representing two distinct
macroscopic groups with opposing opinions. This initial configuration
enables us to investigate the opinion dynamics in the presence of
the barrier, with particular focus on the transition from a polarized
state (characterized by the coexistence of both opinions) to a consensus
state (where only a single opinion exists).

\subsection{Barrier-induced stalemate-consensus transition\protect\label{subsec:Barrier-induced-stalemate-concen}}

In our system, the absence of a barrier ($B=0$) enables initially
separated particles to diffuse freely across the intergroup boundary.
Interfacial particles undergo frequent opinion flipping due to majority-rule
dynamics, resulting in reorganization of the opinion domain boundary.
This process, combined with the ergodicity inherent to random walk
dynamics, drives the system irreversibly to an absorbing consensus
state (all particles with opinion $+1$ or $-1$). In contrast, complete
isolation ($B=L$) prevents both particle exchange and opinion interaction,
indefinitely stabilizing a stalemate state with coexisting opinions.
Between the extremes, we can predict a nonequilibrium phase transition
governed by a critical opening size $O_{c}$ that sharply separates
the consensus and stalemate regimes, analogous to a percolation threshold
in spatial systems (note that $O\equiv L-B$){[}cf. Fig.~\ref{fig:3}{]}.

Our single-particle analysis reveals the origin of the critical opening
size $O_{c}$. The barrier restricts opinion flipping primarily to
particles near the opening during the initial evolution stage, making
these interfacial dynamics crucial for opinion propagation across
parties. Under the random walk assumption, the particle density remains
approximately uniform throughout the system. Besides, each side of
the barrier is basically dominated by a single opinion group in the
initial phrase. These two approximations allow us to relate the local
ratio of opposing opinions to the relative field-of-view areas, which
is partly blocked by the opening {[}cf. Fig.~\ref{fig:1}(b1){]}.
This motivates our definition of an environmental change factor:
\begin{equation}
\gamma=\frac{A_{R}}{A_{L}}\frac{A_{L'}}{A_{R'}}.
\end{equation}

The ratio $A_{R}/A_{L}$ quantifies the instantaneous opinion environment,
representing the relative prevalence of opinion $-1$ versus $+1$
in a given particle's field of view at time $t$. When the particle
crosses the opening at $t+\Delta t$, this ratio inverts to $A_{L'}/A_{R'}$,
reflecting the reversal of its local opinion landscape. The parameter
$\gamma$ combines these ratios and thus characterizes the abruptness
of environmental switch during crossing opening. Under an additional
assumption that particle positions at $t$ and $t+\Delta t$ are mirror-symmetric
about the opening, $\gamma$ simplifies to:
\begin{equation}
\gamma=\frac{A_{R}}{A_{L}}\frac{A_{L'}}{A_{R'}}=\frac{A_{R}}{A_{L}}\frac{A_{R}}{A_{L}}=\left(\frac{A_{R}}{A_{L}}\right)^{2}.
\end{equation}

The parameter $\gamma$ quantifies the stability of opinions during
crossing the opening. Specifically, $\gamma\thickapprox0$ indicates
a drastic switch in the opinion environment, where a particle transitions
from predominantly encountering similar opinions ($t$) to being overwhelmed
by opposing opinions ($t+\Delta t$). This transition facilitates
the particle's rapid assimilation into the opposing opinion group.
Conversely, $\gamma\thickapprox1$ corresponds to a balanced opinion
switch that favors persistent opinion transmission {[}cf. Fig.~\ref{fig:1}(b1,
b2){]}. Since $\gamma$ is monotonically modulated by the opening
size $O$, increasing $O$ drives the system through the stalemate-to-consensus
transition. In other words, a critical opening size $O_{c}$ corresponds
to a critical environmental change factor $\gamma_{c}$.

If view field radius $R$ is increased, maintaining a constant $\gamma$
(i.e., at a certain critical $\gamma_{c}$) requires a corresponding
increase in the opening size. This relationship can be estimated through
simple geometric analysis and is also partially demonstrated by our
simulation results {[}cf. Fig.~\ref{fig:4}(a){]}. Consequently,
a larger view field radius $R$ leads to a larger critical opening
size. Our numerical results indeed show a monotonic increase in the
critical opening size with $R$ {[}cf. Fig.~\ref{fig:1}(c){]}. According
to the trend in Fig.~\ref{fig:1}(c), a much smaller $R$ may result
in a critical opening size close to zero. However, in such a highly
localized view field, the majority rule loses its practical significance,
as the neighbors considered within this limited scope do not constitute
a true \textquotedbl majority\textquotedbl .

In Fig.~\ref{fig:1}(b2), one can observe that the upper bound of
$\gamma$ remains below $1$ in the single-particle picture, regardless
of how much the opening is enlarged. This observation could lead to
the conclusion that a particle will always immediately flip its opinion
upon crossing the opening, thereby precluding the possibility of opinion
transmission. However, this argument overlooks the role of local particle
density fluctuations, which extend beyond the mean-field approximation.
These fluctuations can bridge the crucial gap of $\gamma$ from $\leq1$
to $>1$. Our simulation results provide evidence for this effect.
Even under a relatively small opening $O=0.25L$ (which is covered
by the interval where the consensus probability$P_{\mathrm{co}}$sharply
change), a small but significant proportion of $\gamma$ values are
found to be above $1$, as shown in Fig.~\ref{fig:4}. This finding
indicates that local density fluctuations can facilitate opinion transmission
by allowing $\gamma$ to reach values above $1$, compensating the
limitations in the single-particle picture.

\subsection{Influences of mobility and density\protect\label{subsec:Mobility-modulated-consensus-dyn}}

\begin{figure}[h]
\begin{centering}
\includegraphics[width=1\columnwidth]{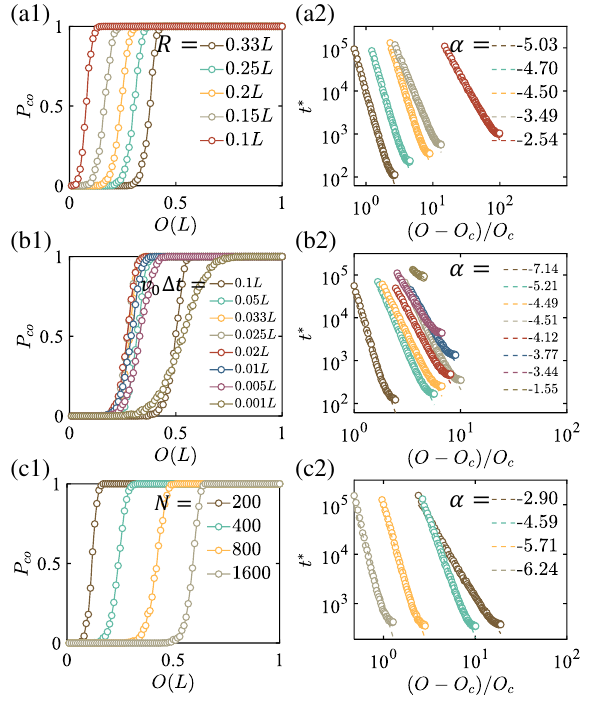}
\par\end{centering}
\caption{\protect\label{fig:3}The scaling behavior near the critical opening
size. The consensus probability $P_{\mathrm{co}}$ (calculated from
$10^{3}$ independent trajectories) is shown for different system
parameters: (a1) view field radius, (b1) particle velocity, and (c1)
particle number. The sharp transitions in P co\LyXZeroWidthSpace{}
reveal the presence of a critical opening size $O_{c}$. In log-log
scale, we plot the consensus time $t^{*}$ near the critical point
and extract the critical exponent $\alpha$ for different (a2) view
field radius, (b2) particle velocities, and (c2) particle numbers.
To avoid misleading consensus times due to the maximal simulation
time limit, we select the $O$ interval where $P_{\mathrm{co}}=1$
throughout for the fitting.}
\end{figure}

\begin{figure}[p]
\begin{centering}
\includegraphics[width=1\columnwidth]{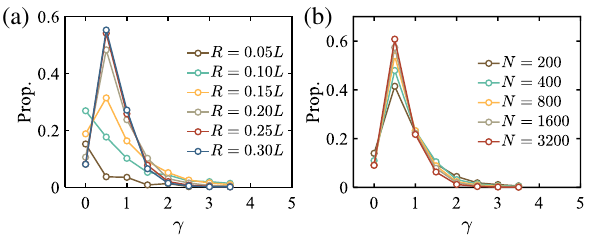}
\par\end{centering}
\caption{\protect\label{fig:4}Numerical results showing the distribution of
the opinion environment factor $\gamma$. (a) As the view field radius
$R$ increases, the proportion of $\gamma$ values below $1$ increases.(b)
Similarly, as the number of particles $N$ increases, the proportion
of $\gamma$ values below $1$ also increases. In (a, b), we fix the
opening size $O=0.25L$.}
\end{figure}

Opinions are embedded in each particle, and thus faster particle movement
is equivalent to faster opinion dissemination. However, when we numerically
tune $v_{0}\Delta t$ from $0.001L$ to $0.1L$, the corresponding
critical opening size $O$ exhibits an interesting V-shaped change,
reaching its minimum around $v_{0}\Delta t=0.025L$ {[}cf. Fig.~\ref{fig:2}(a1){]}.
This behavior can be considered from two aspects. For the regions
where $O_{c}$ increases monotonically (from $0.025L$ to $0.1L$),
the particle velocity increases from a moderate value to an excessively
large one. When the velocity is too high, a particle at time $t+\Delta t$
will enter a much deeper region dominated by the opposing opinion--then
it is isolated by its partners. So it tends to immediately flip its
opinion, which prevents opinion propagation and thus we can see a
higher $O_{c}$ to be an compensation. On the other hand, when the
velocity drops from a moderate to an extremely low value (from $0.025L$
to $0.001L$), a particle will linger longer in the small area near
the opening where both opinions coexist. This prolonged exposure increases
the probability of opinion reversal for a particle, preventing enough
particles from keeping their original opinions after crossing to the
opposite side of the opening. To compensate, the opening must widen
to allow more particles to freely traverse to the other side. Accordingly,
we observe an increase in the critical opening size $O_{c}$.

As discussed in Sec.~\ref{subsec:Barrier-induced-stalemate-concen},
local particle density fluctuations that can lift the opinion environment
factor to $\gamma>1$ are crucial for opinion spread. We can observe
that increasing particle density (equivalent to particle number $N$
for a system of given size) results in a monotonic increase in $O_{c}$
{[}cf. Fig.~\ref{fig:2}(b1){]}. The basic principle of statistical
physics tells us: As particle number $N\rightarrow\infty$, relative
density fluctuations are suppressed. From the perspective of the single-particle
picture, this suppression makes it less likely for $\gamma$ to reach
values high enough to allow a particle to experience a smooth opinion
switch. Therefore, the system needs a larger opening size to reach
consensus. In simulations, we find that the proportion of $\gamma<1$
monotonically increases with particle density as predicted {[}cf.
Fig.~\ref{fig:4}(b){]}.

Critical exponent characterizes how physical quantities depend on
system parameters near a critical point. In our system, it is not
necessary to completely close the opening to prevent opinion merging.
Instead, below a certain opening size $O(>1)$, the consensus probability
$P_{\mathrm{co}}$ undergoes an abrupt drop to zero, indicating the
existence of a barrier-induced opinion transition {[}cf. Fig.~\ref{fig:3}(a1,
b1, c1){]}. Technically, we identify the critical opening size $O_{c}$
as the largest value of $O$ at which $P_{\mathrm{co}}=0$ still exists.
In log-log scale {[}cf. Fig.~\ref{fig:3}(a2, b2, c2){]}, we show
the scaling law of the consensus time $t^{*}$, with the critical
exponent strongly modulated by the system parameters. Through extrapolation,
we can observe the divergence of the consensus time at $O\approx O_{c}$,
which excludes the possibility that the failure to reach a consensus
state in the system is due to limitations in the maximal simulation
time.

\section{Conclusion and outlook\protect\label{sec:Conclusion}}

In this work, we have introduced a minimal model to investigate the
interplay between agent mobility and physical barriers in shaping
collective opinion dynamics. Focusing on conditions under which initially
segregated clusters of agents with opposing opinions can reach consensus,
we find that the system undergoes a transition from a stalemate to
a consensus state at a critical barrier opening size. At this critical
point, the relaxation time required to reach consensus from an initial
stalemate diverges according to a power-law scaling. Beyond establishing
this phase transition, our findings suggest a broader framework for
exploring opinion dynamics in spatially heterogeneous environments.
The present model---where the opinion degrees of freedom do not influence
the translational dynamics---also invites future extensions that
incorporate social or information-driven interactions through opinion-dependent
potentials. We believe this study will stimulate further investigations
into the dynamics of mobile opinion agents in the presence of realistic
spatial constraints, contributing to both the theoretical foundation
and practical modeling of sociophysical systems.\textcolor{black}{}
\begin{acknowledgments}
This work is supported by the NKRDPC (Grant No.~2022YFA1405304),
NSFC (Grant Nos.~12075090, 12275089, and 12475036), and Guangdong
Basic and Applied Basic Research Foundation (Grant Nos.~2022A1515010449,
2023A1515012800, and 2024A1515012575), and Guangdong Provincial Key
Laboratory (Grant No.~2020B1212060066).
\end{acknowledgments}

\bibliographystyle{apsrev4-2}
\bibliography{ref}

\end{document}